\documentclass[twocolumn,showpacs,preprintnumbers,amsmath,amssymb,APSl,nofootinbib,superscriptaddress]{revtex4-1}
\usepackage{dcolumn}
\usepackage{bm}
\usepackage{ifpdf}
\usepackage{hyperref}
\usepackage{dcolumn}
\usepackage{bm}
\usepackage[spanish,english]{babel}
\usepackage{amsfonts}
\usepackage{amssymb}
\usepackage{graphicx}
\usepackage[latin1]{inputenc}

\newcommand{\be}{\begin{equation}}
\newcommand{\en}{\end{equation}}
\newcommand{\bea}{\begin{eqnarray}}
\newcommand{\ena}{\end{eqnarray}}
\newcommand{\bes}{\begin{subequations}}
\newcommand{\ees}{\end{subequations}}
\begin{document}
\title{Accelerated observers and the notion of singular spacetime}
\author{Gonzalo J. Olmo} \email{gonzalo.olmo@uv.es}
\affiliation{Departamento de F\'{i}sica Te\'{o}rica and IFIC, Centro Mixto Universidad de
Valencia - CSIC. Universidad de Valencia, Burjassot-46100, Valencia, Spain.}
\affiliation{Departamento de F\'isica, Universidade Federal da
Para\'\i ba, 58051-900 Jo\~ao Pessoa, Para\'\i ba, Brazil.}
\author{Diego Rubiera-Garcia} \email{drgarcia@fc.ul.pt}
\affiliation{Instituto de Astrof\'{\i}sica e Ci\^{e}ncias do Espa\c{c}o, Faculdade de
Ci\^encias da Universidade de Lisboa, Edif\'{\i}cio C8, Campo Grande,
P-1749-016 Lisbon, Portugal.}
\author{Antonio Sanchez-Puente} \email{sanchez.puente.antonio@gmail.com}
\affiliation{Departamento de F\'{i}sica Te\'{o}rica and IFIC, Centro Mixto Universidad de
Valencia - CSIC. Universidad de Valencia, Burjassot-46100, Valencia, Spain.}
\date{\today}
\begin{abstract}
Geodesic completeness is typically regarded as a basic criterion to determine whether a given spacetime is regular or singular. However, the principle of general covariance does not privilege any family of observers over the others and, therefore, observers with arbitrary motions should be able to provide a complete physical description of the world. This suggests that in a regular spacetime, all physically acceptable observers should have complete paths. In this work we explore this idea by studying the motion of accelerated observers in spherically symmetric spacetimes and illustrate it by considering two geodesically complete black hole spacetimes recently described in the literature. We show that for bound and locally unbound accelerations, the paths of accelerated test particles are complete, providing further support to the regularity of such spacetimes.
\end{abstract}
\pacs{04.20.Dw, 04.50.Kd,04.70.Bw}
\maketitle
\section{Introduction}

Understanding and resolving spacetime singularities became a major goal of any theory of the gravitational field after the establishment of the singularity theorems by Penrose \cite{Penrose1}, Hawking \cite{Hawking}, Carter \cite{Carter}, and others (see \cite{Senovilla1} for a pedagogical discussion). Such theorems are formally based on the notion of \emph{geodesic completeness}, i.e., whether any null or timelike geodesic can be extended to arbitrarily large values of its affine parameter. Moreover, they make use of reasonable assumptions upon the global and causal properties of the geometry, namely, the (null) timelike congruence condition, global hyperbolicity, and the formation of trapped surfaces. This way, they prove that black holes emerging out of gravitational collapse unavoidably contain (at least one) such incomplete geodesic \cite{Wald} (see, however, \cite{Senovilla2}). As null and timelike geodesics are associated to the propagation of information and the free-falling motion of physical objects, respectively, the presence of any of such incomplete curve implies the breakdown of predictability and causality of our physical theories.

As gravity is a matter of geometry, intuitively one would like to have a more direct connection between spacetime singularities and geometrical objects built out of the curvature tensor. In this sense, to avoid introducing artificial singularities in the metric or curvature tensors that can be removed with a choice of basis or frame, working with scalar objects such as curvature invariants is strongly preferred. This way, one would be tempted to blame arbitrary large tidal forces as some spacetime region is approached by a geodesic curve as the underlying reason for its incompleteness, which has shaped numerous attempts in the literature to overcome this difficulty via the bound of curvature scalars \cite{boundedscalars0,BS2}. This intuitive idea was first formalized by Ellis and Schmidt \cite{ES} and latter given rigorous mathematical support by Tipler \cite{Tipler1}, Clarke and Krolak \cite{CK1} and Nolan \cite{Nolan}, among others, by idealizing any physical (extended) object as a congruence of geodesics. This way, according to the classification introduced by Ellis and Schmidt, a \emph{strong} spacetime singularity would crush the observer to zero volume (otherwise, the ``singularity" would be \emph{weak}) as the problematic region is approached. It should be pointed out, however, that such scalar curvature singularities cannot occur unless they are reachable by some geodesic. In this sense, both the Schwarzschild and Reissner-Nordstr\"om solutions of General Relativity (GR), which are strong curvature singularities, can indeed be reached by some geodesics in finite affine time, with no possibility of further extension beyond that point.

Nonetheless, even a spacetime that is both null and timelike geodesically complete and which, in addition, is not affected by arbitrarily large tidal forces, can still display pathological behaviors. This was noted by Geroch in Ref.\cite{Geroch68} (see also \cite{Tod94,GL93} for additional examples), by explicitly showing that there exist spacetimes which contain timelike curves with bounded acceleration and finite proper length\footnote{Conversely, very recently it has been also found that there may exist spacetimes with non-scalar polynomial singularities but free of pathologies as seen from some families of accelerated observers \cite{Burko}.}. This way, a rocket with a finite amount of fuel could meet an incomplete path in finite proper time. As the spacetime is geodesically complete (and inextensible, by assumption in the singularity theorems), there are no further points in the manifold that could be occupied by such a rocket and, by virtue of the principle of general covariance, which does not privilege any family of observers over any other, one would be lead to regard such a spacetime as singular. In agreement with this, a minimum condition for regularity for a given spacetime should extend the geodesically complete requirement to a ``b-complete" one, i.e., that all curves (being geodesic or not) must be complete \cite{Senovilla98}\footnote{A word on caution comes here, as considering arbitrary curves to discuss singularity structure may allow to locally extend some spacetimes, as shown by Beem for the case of Minkowski \cite{Beem80}.}.

The motion of physical observers with acceleration in the context of relativistic field theories has been considered in the literature \cite{Rin}, though only recently in-depth analyses in curved spacetimes have become available. In particular, in Ref.\cite{SF16a} Scarr and Friedman, based on previous notions and work by de la Fuente and Romero \cite{FR15a}, developed a mathematical implementation of the \emph{Frenet-Serret} basis to investigate the modifications of the geodesic equation when otherwise free-falling timelike observers are subject to a uniform acceleration. In addition, they applied this framework to the Schwarzschild geometry of GR, showing that uniformly accelerated observers with finite proper time do exist on it.

The aim of the present paper is twofold. First, to extend the analysis of Scarr and Friedman to any static, spherically symmetric geometry without the restriction of uniform acceleration or radial motions. Our main result will be the obtention of a master equation which accounts for the rate of change of the mechanical energy (per unit mass) of a particle under the action of an acceleration. Second, we will apply this result to the case of two families of geodesically complete black hole geometries recently found in the literature. These geometries, supported by electromagnetic fields and anisotropic fluids satisfying standard energy conditions, reduce to the Reissner-Nordstr\"om solution for far distances but strongly deviate from it close to the central region, where the GR point-like structure is replaced by a wormhole. The nonsingular character of the first family, obtained in the context of Born-Infeld gravity extensions of GR \cite{ors15b} (see \cite{blor17a} for a recent review on such theories), has been established according to the fact that i) it is (null and timelike) geodesically complete for all the spectrum of charge and mass \cite{ors15a}, ii) tidal forces do not unavoidably destroy extended bodies as the central region (where the curvature may be divergent) is crossed \cite{ors16a}, and iii) the scattering of waves off the central wormhole region is well defined \cite{ors15b}. For the second family, obtained within the context of $f(R)$ theories of gravity, the wormhole structure lies on the future (or past) causal boundary and cannot be reached in finite affine time by null geodesics. We shall show that, in both cases,  observers with reasonable (bounded or locally unbounded) accelerations follow complete paths (infinite proper length) and, thus, these geometries retain their nonsingular character according to these criteria.

This paper is organized as follows: in Sec.\ref{sec:equations} we introduce the main elements for the analysis of accelerated observers in spherically symmetric spacetimes, which allow us to obtain a master equation describing both radial and with angular momentum motions. In Sec.\ref{sec:IV} we make use of this equation and consider the accelerated motion of an electrically charged test particle in the context of two geometries which meet the geodesic completeness via the two different mechanisms above. Finally, in Sec.\ref{sec:V} we discuss the implications of our results for the notion of singular spacetimes beyond GR and some perspectives for future research.

\section{Accelerated observers in spherically symmetric spacetimes}\label{sec:equations}

\subsection{Brief review of Frenet-Serret basis}  \label{sec:II}

In a manifold equipped with a metric and an affine connection, one can compute covariant derivatives of a vector field $Z^{\mu}$ along a curve $\gamma(s)$ with unitary tangent vector $u^{\mu}=dx^{\mu}/ds$ as
\begin{equation}
\frac{DZ^{\mu}}{ds}=\frac{d Z^{\mu}}{ds} + \Gamma^{\mu}_{\rho \sigma} Z^{\rho} u^{\sigma} \ ,
\end{equation}
where $s$ is the affine parameter (the proper time for a timelike observer), and $\Gamma_{\mu\nu}^{\lambda}$ are the components of the affine connection, which we assume to be metric compatible unless explicitly stated. In general, the unitary tangent vector to the timelike curve $x^{\mu}(s)$ will not be parallel transported along it, but will change as
\begin{equation} \label{eq:acc}
\frac{Du^{\mu}}{ds}\equiv u^{\mu}\nabla_{\mu}u^{\nu}=a^{\nu} \ ,
\end{equation}
where $a^{\mu}$ is the acceleration vector. Since $u^{\mu}$ is normalized, the above equation implies the orthogonality of these two vectors, $a^{\mu}u^{\nu}g_{\mu\nu}=0$, so if the four velocity is timelike then the four acceleration is spacelike. The equations satisfied by an accelerated timelike observer can be mathematically implemented using a Frenet-Serret frame \cite{SF16b}, which is an orthonormal basis to the tangent space at $x^{\mu}(s)$ written as $\{\lambda_{(0)},\lambda_{(1)},\lambda_{(2)},\lambda_{(3)}\}$, where each component of the basis is some function of the affine parameter, $\lambda_{(a)} \equiv \lambda_{(a)}(s)$, and such that $\lambda^\mu_{(a)}\lambda^\nu_{(b)} g_{\mu\nu}=\eta_{ab}$. If we denote $u^\mu$ as $\lambda^\mu_{(0)}$, then $a^\mu$ can be written as $a^\mu=k(s)\lambda^\mu_{(1)}$, with $k(s)$ denoting the norm of the acceleration vector (also known as {\it curvature} of the curve). Since the $\{\lambda_{(a)}\}$ basis is orthonormal, the $\lambda_{(2)}$ and $\lambda_{(3)}$ elements can be obtained using a Gram-Schmidt procedure (we refer the reader to \cite{SF16b} for full details of this derivation) such that
\begin{equation} \label{eq:FS}
\frac{D\lambda_{(a)}}{ds}=\lambda_{(b)}{A^{(b)}}_{(a)} \ ,
\end{equation}
where $A\equiv {A^{(b)}}_{(a)}$ is a $4 \times 4$ matrix of the form
\bea \label{eq:Amatrix}
{A^{(b)}}_{(a)}=
\left(
\begin{array}{cccc}
0 &  k(s) & 0 & 0\\
k(s) & 0 & \tau_1(s) & 0  \\
0 & -\tau_1(s) & 0 & \tau_2(s)  \\
0 & 0 & -\tau_2(s) & 0  \\
\end{array}
\right)\ ,  \label{eq:em}
\ena
where the functions $\{k(s),\tau_1(s),\tau_2(s)\}$ are called the curvature, first torsion and second torsion of the curve $\gamma(s)$, respectively\footnote{Note that we use the tangent space metric $\eta_{ab}=\text{diag}[-1,1,1,1]$, which has opposite signature to that employed in \cite{SF16b}}. The physical interpretation of $k(s)$ is that of the linear acceleration experienced by an observer in the direction of $\lambda_{(1)}$. As for the torsions $\tau_1(s)$ and $\tau_2(s)$, when they are non-vanishing an observer experiences a rotational acceleration of magnitude $\sqrt{\tau_1^2+\tau_2^2}$ along an axis determined by the vector $\omega=\tau_2 \lambda_{(1)}+\tau_1\lambda_{(3)}$. Note that the \emph{acceleration} matrix $A$ does not change upon a transformation of coordinates, so parenthesis in the indices have been introduced to stress this coordinate-free character.

The Frenet-Serret equations (\ref{eq:FS}) provide a natural generalization of the well known geodesic equation for free-falling observers. This can be checked after setting $k=\tau_1=\tau_2=0$ (i.e. zero acceleration, $A=0$), and considering a tangent vector field $\lambda_{(0)}=u^{\mu}(s) \equiv dx^{\mu}/ds$ the field equations (\ref{eq:FS})  become (for $k=0$)
\begin{equation} \label{eq:geoeqgr}
\frac{du^{\mu}}{ds}+\Gamma^{\mu}_{\rho\sigma} u^{\rho}(s)u^{\sigma}(s)=0 \ .
\end{equation}
This  set of second-order differential equations has an unique solution once initial conditions $x^{\mu}(0)$ and $u^{\mu}(0)$ are given.

\subsection{Radial accelerated motion}  \label{sec:III}

In this section we shall particularize the Frenet-Serret equations (\ref{eq:FS}) to the case of spherically symmetric and static spacetimes, for which an analytical solution will be possible. Let us thus take, for convenience, the following notation: $ \{ \lambda_{(0)}=\hat u, \lambda_{(1)}=\hat e, \lambda_{(2)}=\hat{\theta}, \lambda_{(3)}=\hat{\varphi} \}$ to separate the angular part from the temporal and radial ones. In addition, we will restrict the analysis to \emph{linear} accelerations only, this is, with no acceleration in the angular directions ($\tau_1=\tau_2=0$). The Frenet-Serret equations (\ref{eq:FS}) in this case read explicitly:
\begin{eqnarray}
\frac{du^{\mu}}{ds} + \Gamma_{\alpha\beta}^{\mu}u^{\alpha}u^{\beta}&=&k e^{\mu} \label{eq:FSu} \\
\frac{de^{\mu}}{ds} + \Gamma_{\alpha\beta}^{\mu}u^{\alpha}e^{\beta}&=&k u^{\mu} \label{eq:FSe} \\
\frac{d\hat{\theta}^{\mu}}{ds} + \Gamma_{\alpha\beta}^{\mu}u^{\alpha}\hat{\theta}^{\beta}&=&0 \label{eq:FSt} \\
\frac{d\hat{\varphi}^{\mu}}{ds} + \Gamma_{\alpha\beta}^{\mu}u^{\alpha}\hat{\varphi}^{\beta}&=&0\ . \label{eq:FSv}
\end{eqnarray}
The first two equations (\ref{eq:FSu}) and (\ref{eq:FSe}) correspond to the accelerated motion while (\ref{eq:FSt}) and (\ref{eq:FSv}) simply express parallel transport along the angular sector. Now, let us consider a static, spherically symmetric line element of the form\footnote{For the sake of the discussion below, in this line element we introduce three independent functions, though one of them can always be eliminated in favour of the other two via a change of coordinates. Later on we will make use of this freedom in fixing the gauge for our advantage.}
\begin{equation} \label{eq:linele}
ds^2=-A(x)dt^2 +\frac{1}{B(x)}dx^2+r^2(x)d\Omega^2 \ ,
\end{equation}
where the functions $A(x)$, $B(x)$ and $r^2(x)$ specify the geometry. The components of the connection appearing in the Frenet-Serret equations are easily computed as the Christoffel symbols of the metric (\ref{eq:linele}). This way, the relevant components of the Levi-Civita connection for our problem read
\begin{equation} \label{eq:LC}
 \Gamma^x_{tt}= \frac{(\partial_x A)}{2B} \hspace{0.1cm}; \hspace{0.1cm}
 \Gamma^x_{xx}= \frac{(\partial_x B)}{2B} \hspace{0.1cm}; \hspace{0.1cm}
 \Gamma^x_{\theta \theta} = - \frac{r (\partial_x r)}{B} \ .
\end{equation}
Let us now focus on purely radial motions of the timelike observer (the case with angular momentum will be separately treated in Sec. \ref{Sec:AM} below). This choice amounts to $\{u^{\theta}=u^{\varphi}=e^{\theta}=e^{\varphi}=0\}$. Now,  writing $u^{\mu}=(u^t,u^x,0,0)$ and $e^{\mu}=(e^t,e^x,0,0)$, we can explicitly separate each of the Eqs.(\ref{eq:FSu}) and (\ref{eq:FSe}) into their non-vanishing $t$ and $x$ components as
\begin{eqnarray}
\frac{du^t}{ds}&+&\frac{A_x}{A} u^tu^x=k e^t \label{eq:uteq} \\
\frac{du^x}{ds}&+&\frac{BA_x}{2}(u^t)^2-\frac{B_x}{2B}(u^x)^2=k e^x \label{eq:uxeq} \\
\frac{de^t}{ds}&+&\frac{A_x}{2A}(e^tu^x+e^xu^t)=k u^t \label{eq:eteq} \\
\frac{de^x}{ds}&+&\frac{BA_x}{2}u^te^t-\frac{B_x}{2B} u^xe^x=k u^x \ ,  \label{eq:exeq}
\end{eqnarray}
where from now on we shall use the short-hand notation $A_x \equiv dA/dx$ and $B_x \equiv dB/dx$. This system of equations is further constrained using the orthonormality of  $u^{\mu}$ and $e^{\mu}$ as well as their timelike and spacelike character, respectively, i.e.:
\begin{equation}
u^{\mu}e^{\nu}g_{\mu\nu}=0\hspace{0.1cm};\hspace{0.1cm}u^{\mu}u^{\nu}g_{\mu\nu}=-1\hspace{0.1cm};\hspace{0.1cm}e^{\mu}e^{\nu}g_{\mu\nu}=+1 \ .
\end{equation}
In this case, from the line element (\ref{eq:linele}), these equations read
\begin{eqnarray}
-Au^te^t&+&\frac{1}{B}u^xe^x=0 \label{eq:cons1} \\
-A(u^t)^2&+&\frac{1}{B}(u^x)^2=-1 \label{eq:cons2}  \\
-A(e^t)^2&+&\frac{1}{B}(e^x)^2=+1 \label{eq:cons3}  \ ,
\end{eqnarray}
which impose constraints upon the functions $\{u^t,u^x,e^t,e^x\}$ in such a way that only one of the equations (\ref{eq:uteq}), (\ref{eq:uxeq}), (\ref{eq:eteq}) and (\ref{eq:exeq}) remains independent. For convenience, let us choose $u^x$ as the independent object, so from (\ref{eq:cons1}), (\ref{eq:cons2}) and (\ref{eq:cons3}) one gets
\begin{equation} \label{eq:cons}
(u^t)^2=\frac{1}{A}\left[1+\frac{(u^x)^2}{B} \right];  (e^t)^2=\frac{(u^x)^2}{AB}  ;  (e^x)^2=B+(u^x)^2 \ ,
\end{equation}
and, consequently, Eq.(\ref{eq:uxeq}) reads now
\begin{equation} \label{eq:duxds}
\frac{du^x}{ds}+\frac{BA_x}{2A} + (u^x)^2\left[\frac{A_x}{2A}-\frac{B_x}{2B} \right] =k(s) \sqrt{B+(u^x)^2} \ ,
\end{equation}
Multiplying this equation by $u^x=dx/ds$ and manipulating it, we get
\begin{equation} \label{eq:duxds2}
\frac{B}{A}\frac{d}{ds}\left(\frac{A}{B}(u^x)^2+A\right) =2k(s)u^x \sqrt{B+(u^x)^2} \ ,
\end{equation}
which can be written in the more suggestive form
\begin{equation} \label{eq:duxds3}
d\left(\sqrt{\frac{A}{B}(u^x)^2+A} \right)=k(s)\sqrt{\frac{A}{B}}dx \ ,
\end{equation}
where the affine parameter can be seen as $s=s(x)$. The formal integration of this equation thus leads to
\begin{equation} \label{eq:duxds4}
\frac{A}{B}(u^x)^2+A=\left(E+\int^x_{x_0} dx' k(s)\sqrt{\frac{A}{B}}\right)^2 \ ,
\end{equation}
where $E$ is an integration constant whose meaning shall be clarified at once. It is worth noting that with the change of coordinates defined by $dy^2=(A/B)dx^2$,
the line element (\ref{eq:linele}) becomes
\begin{equation} \label{eq:ley}
ds^2 = - A(y) dt^2 + \frac{1}{A(y)} dy^2 + r^2(y) d \Omega^2 \ ,
\end{equation}
and Eq.(\ref{eq:duxds4}) turns into the simpler form
\begin{equation} \label{eq:duxds5}
(u^y)^2+A(y)=\left(E+\int^y_{y_0} dy' k(s)\right)^2 \ ,
\end{equation}
where $u^y\equiv dy/ds$. When the radial acceleration vanishes, $k(s)=0$, then Eq.(\ref{eq:duxds5}) yields\begin{equation} \label{eq:Econservation}
E^2=(u^y)^2+A(y)\ ,
\end{equation}
which simply expresses the conservation of the total mechanical energy per unit mass, $E$, in absence of acceleration. This follows from the fact that, using the first of Eqs.(\ref{eq:cons}), the energy can be expressed as $E^2=A^2 (u^t)^2=A^2(dt/ds)^2$, which is a well known conserved quantity in spherically symmetric, static systems \cite{OlmoBook}.

The above equation (\ref{eq:duxds5}) is the main result of this section. It illustrates how the mechanical energy (per unit mass) of the particle changes due to the action of a radial acceleration of amplitude $k(s)$. For non-accelerated test particles, the mechanical energy (per unit mass) $E$ remains constant, as it should be for geodesic motion.

For completeness, we note that the vectors $\hat{\theta}^\mu$ and $\hat{\varphi}^\mu$ that complete the Frenet-Serret basis can be written in this case as  $\hat{\theta}^\mu=(0,0,1/r(y),0)$ and $\hat{\varphi}^\mu=(0,0,0,1/(r(y)\sin\theta_0))$, where $\theta_0$ is a constant angle that specifies the plane on which the motion takes place. It is straightforward to verify that all the orthonormality relations are satisfied for these basis vectors.

\subsection{Accelerated motion with angular momentum} \label{Sec:AM}

The analysis presented above can be extended to radially accelerated observers with constant angular momentum. For simplicity we will now consider directly the line element (\ref{eq:ley}).  The first thing to note is that the angular momentum per unit mass, $L \equiv r^2(y) \sin^2 \theta d\varphi/ds$, is a conserved quantity due to spherical symmetry and that we can, in addition, rotate the system of coordinates for the movement to take place in the equatorial plane ($\theta_0=\pi/2$) of the symmetric $2$-spheres without loss of generality. Therefore, since $u^\varphi=L/r^2$, the normalization of $u^\mu$  allows us to parameterize the components of this vector in terms of $u^y$ as follows:
\begin{equation}\label{ucomp}
 u^\mu = \left ( \frac{1}{A} \sqrt{ A\left ( 1 + \frac{L^2}{r^2} \right ) + (u^y)^2}, u^y, \frac{L}{r^2},0 \right ) \ .
\end{equation}
Since we are assuming vanishing torsions in the angular directions, the acceleration vector $a^\mu$ has only two components. As a result, it can also be parameterized in terms of $u^y$. Using the normalization of $e^\mu$ and its orthogonality with $u^\mu$, we find
\begin{equation}\label{acomp}
 a^\mu = \frac{k(s)}{\sqrt{1+\frac{L^2}{r^2}}} \left ( \frac{u^y}{A}, \sqrt{A \left ( 1+ \frac{L^2}{r^2} \right ) +(u^y)^2},0,0 \right ) \ .
\end{equation}
We note that $a^\varphi\equiv Du^\varphi/ds$, with $u^\varphi=L/r^2$, is identically zero due to conservation of angular momentum. The equation for the $u^y$ component follows from
\begin{equation}
 \frac{d u^y}{ds} + \Gamma^y_{\mu \nu} u^\mu u^\nu = \frac{k(s)}{\sqrt{1+\frac{L^2}{r^2}}} \sqrt{A \left ( 1+ \frac{L^2}{r^2} \right ) +(u^y)^2} \ .
\end{equation}
Making explicit the connection coefficients, multiplying by $u^y$, and rearranging terms, one ends up with
\begin{equation}
d\left[(u^y)^2+A\left(1+\frac{L^2}{r^2}\right)\right]^{1/2}=\frac{k(s)dy}{\sqrt{1+\frac{L^2}{r^2}}} \ ,
\end{equation}
which can be integrated to obtain
\begin{equation}\label{eq:acceL}
(u^y)^2+A(y)\left(1+\frac{L^2}{r^2(y)}\right)=\left[E+\int^y_{y_0}\frac{k(s)dy'}{\sqrt{1+\frac{L^2}{r^2(y')}}}\right]^2 \ .
\end{equation}
This expression is the natural extension of (\ref{eq:duxds5}) to the case with non-zero angular momentum. It is possible to find an alternative way to reach to this same expression looking for conserved quantities in the unaccelerated case. Given a killing vector $\tau^\alpha=\left ( \partial /\partial t \right )$ and a geodesic with unit tangent vector $u^\alpha$, the energy (per unit mass) $E\equiv-\tau_\alpha u^\alpha$ is conserved. However, if instead of a geodesic we have an accelerated curve, this quantity changes as:
\begin{equation}
 u^\beta \nabla_\beta ( - \tau_\alpha u^\alpha) = - a^\alpha \tau_\alpha
 \end{equation}
Integrating this expression between two points $a$ and $b$ for the geometry (\ref{eq:ley}) we find:
\begin{equation}
 \left. E \right |_a^b = -\int_a^b a^\alpha \tau_\alpha d \lambda = -\int_a^b \frac{a^t g_{tt}}{u^y} d y
\end{equation}
This expression is equivalent to Eq.(\ref{eq:acceL}): the energy at $b$ can be expressed in terms of the radial value and radial component of the unit tangent vector as
\begin{equation}
 E(b) = - u^t g_{tt} = \sqrt{(u^y)^2+A(y)\left(1+\frac{L^2}{r^2(y)}\right)} \ ,
\end{equation}
and substituting $g_{tt}$, $u^y$ and $a^t$ with the value found in Eqs.(\ref{eq:ley}), (\ref{ucomp}) and (\ref{acomp}), we get to Eq.(\ref{eq:acceL}), where the integration constant $E$ is simply the energy at the beginning of the accelerated motion $E(a)$. This way it is straightforward to see that the energy can only take negative values in the regions of spacetime where the killing vector $\tau^\alpha$ becomes space-like, such as behind the event horizon in a Schwarzschild black hole, or inside the ergosphere in a Kerr black hole. Also, if we would like the accelerated curve to ``escape'' from the light-cone, it would need to achieve infinite energy, and consequently, the acceleration needed would also be infinite.

In order to complete the Frenet-Serret basis, it is important to note that the orthogonality condition between $\hat{\varphi}$ and the basis vectors $\hat u$ and $\hat e$ requires that $\hat{\varphi}$ has three non-vanishing components, $\hat{\varphi}=(\varphi^t, \varphi^x,0,\varphi^\varphi)$. After some algebra, one can verify that these components take the form
\begin{eqnarray}
\varphi^t&=& \frac{L^2}{r^2}\frac{\left[(u^y)^2+A(y)\left(1+\frac{L^2}{r^2}\right)\right]}{A^2(y)\left(1+\frac{L^2}{r^2}\right)}\\
\varphi^x&=& \frac{L^2}{r^2}\frac{(u^y)^2}{\left(1+\frac{L^2}{r^2}\right)}\\
\varphi^\varphi&=& \frac{1}{r^2}\left(1+\frac{L^2}{r^2}\right) \ ,
\end{eqnarray}
where we have taken $\sin\theta_0=1$. The vector $\hat\theta$ is the same as in the case $L=0$.

This concludes our analysis of radially accelerated observers with constant angular momentum in the context of static, spherically symmetric geometries. We shall employ next these results to test the nonsingular character of some geometries recently found in the literature.

\section{Accelerated observers and the notion of singular spacetime}  \label{sec:IV}

The main purpose of this section is to deepen into the notion of singular spacetime by exploring if in a geodesically complete geometry accelerated observers could have incomplete paths. As mentioned in the introduction, a spacetime is typically regarded as singular if there exist incomplete geodesics on it. This incompleteness implies that light rays and/or freely falling observers could disappear or come into existence from nowhere, which poses a serious problem for the  predictability of the laws of Physics and even to our ability to make measurements. However, even in a geodesically complete space, there is no objective reason to prefer freely falling observers over accelerated ones in order to determine whether it is singular or not, since both are equally important as far as the possibility of making measurements is concerned. Moreover, the principle of general covariance states that the laws of Physics need not be referred to any special family of observers, and that any non-degenerate choice of coordinates should be physically admissible.  Thus, if the trajectories of freely falling observers were safe (complete paths) but a family of accelerated observers could disappear from the spacetime or be created out of nowhere at some instant of time (incomplete paths), then such a spacetime could be regarded as pathological\footnote{Obviously, an observer in such an accelerated worldline could always modify the thrust of its rocket engine to drop off from that worldline and be safe. }.

Before considering specific examples of accelerated observers, it is necessary to comment on a relevant implicit element of our approach. The equations that describe accelerated observers in Sec. \ref{sec:equations} are an extension of the geodesic equations for neutral test particles. These equations assume that there exists a robust background geometry which is not affected by the presence of the test particle (negligible backreaction). If the energy invested in accelerating the particle is to have very little or no effect on the background geometry, it seems natural to expect that the particle should be able to reach only bounded accelerations, as otherwise the backreaction could be large. One could, however, consider very light particles able to experience huge accelerations and abrupt changes of direction (though maintaining a continuous path). So there is, in principle, a broad spectrum of possibilities.  For this reason, and given the technical difficulties that a detailed and consistent description of arbitrarily accelerated particles could imply, we will consider a somewhat crude approximation and neglect any effect that the test particle could have on the geometry. Furthermore, we will just consider an electrically charged test particle, for which the acceleration felt near the central object is very large, though we will assume it to be bounded in all cases, in agreement with the underlying quantum-mechanical nature of fundamental particles. This choice is justified since the resulting scenario is technically simple and natural. Thus, our model of accelerated particle will be of the form (\ref{eq:acc}) with
\begin{equation}
a^\mu=(Q/m) {F^\mu}_\nu u^\nu \ ,
\end{equation}
being $Q$ and $m$ the charge and mass of the test particle, and ${F^\mu}_\nu$ the Faraday tensor associated to the central object that generates the gravitational and electric fields.

Next we discuss the motion of this charged test particle in two geodesically complete spacetimes with different properties.

\subsection{Born-Infeld gravity with a central electric field} \label{sec:IVA}

The first model that we consider corresponds to an extension of GR dubbed as Eddington-inspired Born-Infeld gravity (EiBI), originally introduced by Deser and Gibbons \cite{DG}, and latter studied by many authors in the context of astrophysics, black hole physics and cosmology \cite{BIapp} (see \cite{blor17a} for a recent review on this class of theories). It is defined by the action
\begin{eqnarray} \label{eq:actionBI}
\mathcal{S}_{EiBI}&=&\frac{1}{\kappa^2 \epsilon} \int d^4x \left[ \sqrt{\vert g_{\mu\nu} + \epsilon R_{\mu\nu} \vert} -\lambda \sqrt{-g} \right] \\
&+& \mathcal{S}_m(g_{\mu\nu},\psi_m) \nonumber \ ,
\end{eqnarray}
with the following definitions: $\kappa^2$ is Newton's constant in suitable units (in GR, $\kappa^2 =8\pi G/c^3$), $\epsilon$ is EiBI parameter with dimensions of length squared, $g$ is the determinant of the spacetime metric $g_{\mu\nu}$, vertical bars denote a determinant, the (symmetrized) Ricci tensor $R_{\mu\nu}=R_{(\mu\nu)}(\Gamma)$ is entirely built out of the affine connection, and $\mathcal{S}_m$ corresponds to the matter action, where $\psi_m$ labels collectively all the matter fields. The physical content of the parameter $\lambda$ comes from taking the limit on which the curvature $R_{(\mu\nu)}$ is much smaller than the scale $1/\epsilon$:
\begin{eqnarray}
\mathcal{S}_{EiBI}&\approx &\frac{1}{2\kappa^2} \int d^4x \sqrt{-g} \Big[R - 2\Lambda_{eff}   \label{eq:actionquad} \\
&+& \frac{\epsilon}{2} \Big(\frac{R^2}{2}-R_{\mu\nu}R^{\mu\nu}\Big)\Big] + \mathcal{O}(\epsilon^2) + \mathcal{S}_m(g_{\mu\nu},\psi_m) \ , \nonumber
\end{eqnarray}
where $\Lambda_{eff}=\frac{\lambda-1}{\epsilon}$ plays the role of an effective cosmological constant. Thus this theory modifies the GR dynamics only in regions of high-curvature (or high-energy density) and, consequently, it is useful in order to investigate new high-energy physics beyond GR.

It should be pointed out that, in order to avoid troubles with higher-order derivative field equations and ghosts, EiBI gravity is typically formulated in a \emph{Palatini} or metric-affine approach, where metric and affine connection are regarded as independent degrees of freedom. Due to this one could wonder whether the connection appearing in the Frenet-Serret equations (\ref{eq:FS}) in this case is given by the independent connection $\Gamma^{\lambda}_{\mu\nu}$ or by the connection associated to the Christoffel symbols of the metric $g_{\mu\nu}$. To answer this question, we note that in the  action (\ref{eq:actionBI}) we have chosen the matter fields not to couple to the connection. In turn this implies, via the field equations, that the independent connection $\Gamma_{\mu\nu}^{\lambda}$ is a nondynamical field which can be completely removed in favor of the matter fields and the metric, and whose only effect is to modify the structure of the geometry $g_{\mu\nu}$, but plays no further role in the geodesic motion (see \cite{ors15a} for more details on this important point). This result guarantees that test particles will follow geodesics of the spacetime metric and thus the weak equivalence principle will be satisfied\footnote{In this sense, should one consider the direct coupling of the matter sector to the connection, then geodesics of the independent connection would be physically meaningful.}. This means that we can safely make use of the Frenet-Serret equations (\ref{eq:FS}) with the Christoffel symbols of the metric $g_{\mu\nu}$ and, therefore, all the analysis carried out in Sec. \ref{sec:III} is valid also in this case.

Having said this, we can safely return to EiBI gravity. Assuming a spherically symmetric (Maxwell) electrovacuum field with Minkowskian asymptotics, $\lambda=1$, the field equations associated to the action (\ref{eq:actionBI}) yield a geometry of the form  (see \cite{ors13} for full details of this derivation)
\begin{eqnarray}
ds^2&=&-A(x)dt^2+\frac{1}{A(x)}\left(\frac{dx}{\sigma_{+}}\right)^2 + r^2(x)d\Omega^2 \label{eq:leP} \\
A(x)&=& \frac{1}{\sigma_+}\left[1-\frac{r_S}{ r  }\frac{(1+\delta_1 G(r))}{\sigma_-^{1/2}}\right] \label{eq:ABI} \\
\delta_1&=& \frac{1}{2r_S}\sqrt{\frac{r_q^3}{l_\epsilon}} \label{eq:delta1} \\
\sigma_\pm&=&1\pm \frac{r_c^4}{r^4(x)} \label{eq:sigma} \\
\left(\frac{dr}{dx}\right)^2&=&\frac{\sigma_{-}}{\sigma_{+}^2} \rightarrow  r^2(x)= \frac{x^2+\sqrt{x^4+4r_c^4}}{2} \label{eq:r(x)} \ ,
\end{eqnarray}
where $r_S=2M_0$ is Schwarzschild radius, $r_q=2Gq^2$ the charge radius, and we have redefined $\epsilon=-2l_{\epsilon}^2$ (and assumed $l_{\epsilon}^2>0$), and the function $G(z)$, with $z=r/r_c$ (where $r_c=\sqrt{r_ql_{\epsilon}}$), satisfies $dG/dz=\sigma_{+}/(z^2\sigma_{-}^{1/2})$, which can be explicitly integrated as
\begin{equation} \label{eq:G(z)}
G(z)=-\frac{1}{\delta_c}+\frac{1}{2}\sqrt{z^4-1}\left[f_{3/4}(z)+f_{7/4}(z)\right] \ ,
\end{equation}
being $f_\lambda(z)={_2}F_1 [\frac{1}{2},\lambda,\frac{3}{2},1-z^4]$  a hypergeometric function, and $\delta_c\approx 0.572069$ a constant needed to ensure agreement between the far and central behaviors. This is a geometry characterized by charge and mass which quickly boils down to the Reissner-Nordstr\"om solution of GR a few $r_c$ units away from the center. However, important departures from the GR solution are found as the region $r=r_c$ is approached. This can be easily seen from the behavior of the radial function in Eq.(\ref{eq:r(x)}), which reaches a minimum at $x=0$ ($r=r_c$ or $z=1$) and bounces off there (see Eq.(\ref{eq:delta1})), representing the throat of a wormhole (see \cite{Visser} for a detailed account of wormhole physics and \cite{Lobo:2017oab} for recent developments).

For our purposes, the most relevant aspect of this solution is the fact that the wormhole is supported by an electric field. Given that for Maxwell's electrodynamics the modulus of the acceleration experienced by our test particle is $k(y)=\frac{Qq}{m r^2(y)}$, the nonzero size of the wormhole implies that $k(y)$ is bounded (unlike in GR, where the field emanates from $r= 0$). Moreover, the presence of this wormhole structure introduces significant deviations in the geodesic behavior as compared to the GR counterparts (the latter containing incomplete radial null geodesics \cite{Chandra}). Indeed, as follows from the detailed analysis carried out in Ref.\cite{ors15a}, this geometry is timelike, null, and space-like geodesically complete for all the spectrum of mass and charge of the solution (encoded in the constant $\delta_1$ in Eq.(\ref{eq:delta1})).

\begin{figure}[h]
\includegraphics[width=0.5\textwidth]{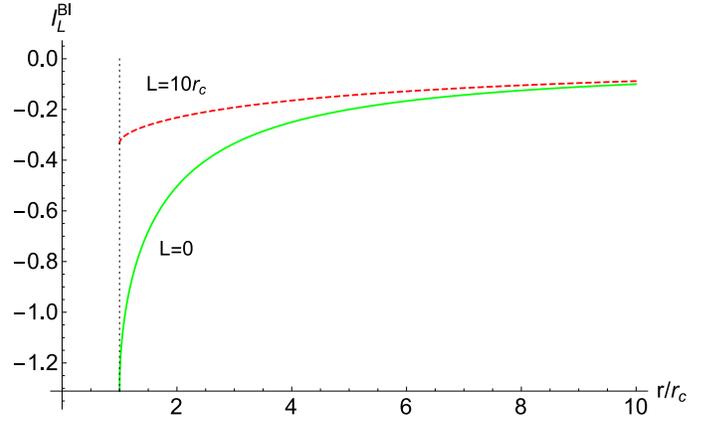}
\caption{ Representation of $(m/Qq)I^{BI}_L(r)$ of Eq.(\ref{eq:IBIL}) as a function of $r/r_c$ for two values of $L$.   }\label{fig:IBIL}
\end{figure}

Let us now consider the behavior of a charged test particle in this geometry by looking at Eq.(\ref{eq:acceL}). Instead of using the radial variable $x$ of Eq.(\ref{eq:leP}), we will introduce the coordinate $dy=dx/\sigma_+$ and make use of the relation $dx=(\sigma_+/\sqrt{\sigma_-})dr$ of Eq.(\ref{eq:r(x)}) so that the integral in that formula turns into
\begin{equation}
I^{BI}_L(r)=\frac{Qq}{m}\int^r \frac{dr}{r^2\sigma_-^{\frac{1}{2}}\sqrt{1+L^2/r^2}} \ .
\end{equation}
This integral admits an exact solution of the form
\begin{equation}\label{eq:IBIL}
I^{BI}_L(r)=-\frac{Qq}{m} \frac{\text{EllipticF}\left[\sin ^{-1}\left(\sqrt{\frac{L^2+r_c^2}{L^2+r^2}}\right),\frac{L^2-r_c^2}{L^2+r_c^2}\right]}{\sqrt{r_c^2+L^2}}\ .
\end{equation}
One easily verifies that for $r\gg r_c$ this expression recovers the GR prediction, $I^{BI}_L(r)\approx -\frac{Qq}{m}\left[\frac{1}{r}-\frac{L^2}{6 r^3}\right]$. Near the wormhole throat, $r=r_c$, the behavior is radically different, turning into
\begin{equation}
I^{BI}_L(r)\approx \frac{Qq}{mr_c}\left( C_0+\sqrt{\frac{r-r_c}{r_c(1+L^2/r_c^2)}}\right) \ ,
\end{equation}
where $C_0\equiv {\text{EllipticK}\left[\frac{2 L^2}{L^2+r_c^2}-1\right]}/{\sqrt{1+L^2/r_c^2}} $ is a constant which goes as $\sim -(r_c/L)\ln(2^{3/2}L/r_c)$ as $L$ grows and tends to $C_0\to -\sqrt{\pi}\Gamma[\frac{5}{4}]/\Gamma[\frac{3}{4}]\approx -1.311$ as $L\to 0$ (see Fig. \ref{fig:IBIL}).

The trajectory of our charged test particle, therefore, is determined by
\begin{equation}\label{eq:beyondGeoBI}
\left(\frac{u^x}{\sigma_+}\right)^2+A(x)\left(1+\frac{L^2}{r^2(x)}\right)=\left(\tilde{E}+I^{BI}_L(r)\right)^2 \ ,
\end{equation}
where $\tilde{E}\equiv E-I^{BI}_L(r_i)$. It is easy to see that in the asymptotic far region, the $1/r$ leading-order terms recover the expected Newtonian behavior
\begin{equation}
E_{NR}=\frac{m (u^x)^2}{2}+\frac{Qq}{r}-\frac{mM}{r} \ ,
\end{equation}
where we have denoted the non-relativistic mechanical energy as $E_{NR}\equiv m(\tilde{E}^2-1)$, and $u^x\equiv dx/d\tau\approx dr/d\tau$.

Since the far away behavior is identical to what happens in GR, we will just focus on the region close to the wormhole, where the electric field intensity reaches its maximum. In this sense,  from Eqs.(\ref{eq:ABI}) and (\ref{eq:G(z)}), expanding $A(x)$ around $x=0$ and using $r-r_c\approx x^2/4r_c$ yields the result
\begin{eqnarray}\label{eq:A_expansion}
\lim_{r\to r_c} A(x) &\approx & \frac{N_q}{2N_c}\frac{\left(\delta _1-\delta _c\right) }{\delta _1 \delta _c }{\frac{r_c}{ |x|} }+\frac{N_c-N_q}{2 N_c} \nonumber \\
&+& \mathcal{O}\left(\sqrt{r-r_c}\right) \ ,
\end{eqnarray}
where $N_q=|q/e|$ is the number of charges and $N_c =\sqrt{2/\alpha_{em}} \simeq 16.55$ (with $\alpha_{em}$ the fine structure constant). This equation implies that the behavior of the function $A(x)$ at the center is controlled by the ratio $\delta_1/\delta_c$, i.e., if $\delta_1>\delta_c$ then $A(x) \rightarrow +\infty$, if $\delta_1<\delta_c$ then $A(x) \rightarrow  -\infty$, while if $\delta_1=\delta_c$ then $A(x)\approx {(N_c-N_q)}/{2 N_c}$  \footnote{Indeed, the ratio $\delta_1/\delta_c$ also determines the number and type of horizons in these geometries, see \cite{ors15a} for details.}. With this result, around $x=0$ Eq.(\ref{eq:beyondGeoBI}) can be written as
\begin{equation}\label{eq:beyondGeoBI}
\left(\frac{dx}{2d\tau}\right)^2+A(x)\left(1+\frac{L^2}{r_c^2}\right)\approx \left(\tilde{E}+\frac{Qq}{m r_c}C_0\right)^2 \ ,
\end{equation}
where $Qq /m r_c\approx-2\alpha_{em} (N_Q N_q/N_c^2)(\lambda_m/r_c)$, being $\lambda_m\equiv \hbar/mc$ the reduced Compton wavelength of the test particle.  It is remarkable that this equation is identical to that for geodesics with the replacement $\tilde{E}\to \tilde{E}+(Qq/mr_c)C_0$.
The fact that the right-hand side of this equation is a constant that simply shifts the energy $\tilde{E}$ of freely falling observers indicates that the resulting trajectories have no pathologies and can be extended to arbitrary values of the affine parameter. Like in the non-accelerated case, see \cite{ors15a} for details, all trajectories in the $\delta_1>\delta_c$ case bounce before reaching the throat, while those with $\delta_1<\delta_c$ can be smoothly extended across the wormhole. When $\delta_1=\delta_c$, it is the number of charges $N_q$ which determines whether there is a bounce or not before reaching the throat.

It is worth noting that depending on the sign of $\tilde{E}$ and of the product $Qq$, the right-hand side of (\ref{eq:beyondGeoBI}) could vanish as the particle approaches the wormhole. The fate of the particle in those cases depends on the properties of the left-hand side. If the function $A(x)>0$ when the right-hand side vanishes, then the particle motion should have experienced a bounce before getting to that point because, otherwise, one would have $(u^x)^2<0$, which is absurd. On the contrary, if $A(x)<0$ at that point, then there is no inconsistency with the positivity of $(u^x)^2$ and the particle could continue its trajectory towards the wormhole to eventually cross it. Therefore, the potential vanishing of the right-hand side does not seem to have any new relevant implications.

Before concluding this subsection, we note that curvature scalars such as $R$, $R_{\mu\nu}R^{\mu\nu}$, $R_{\alpha\beta\mu\nu}R^{\alpha\beta\mu\nu}$, \ldots,  in this model are (provided that $\delta_1 \neq \delta_c$) divergent at the wormhole throat. In spite of this, geodesics are complete and the accelerations produced by the electric field are bounded. Therefore it seems safe to conclude that this geometry is nonsingular.

\subsection{Quadratic $f(R)$ gravity with non-linear electrodynamics and anisotropic fluid} \label{eq:subb}

A different family of nonsingular black holes recently found in the literature arises in an $f(R)$ extension of GR (see \cite{fRr,fR} for motivations and application of such theories in the literature), also formulated in the Palatini approach. The prototypical action in this case is
\begin{equation} \label{eq:actionfR}
\mathcal{S}=\frac{1}{2\kappa^2} \int d^4x \sqrt{-g} \left(R-\alpha R^2 \right)+\mathcal{S}_m(g_{\mu\nu},\psi_m) \ ,
\end{equation}
where $\alpha$ is a parameter with dimensions of length squared and similar definitions as in action (\ref{eq:actionBI}) apply. Taking the matter sector in (\ref{eq:actionfR}) to be either given by a non-linear electromagnetic Born-Infeld field \cite{bcor15} or by some kind of anisotropic fluid satisfying the energy conditions \cite{or15c}, wormhole structures replacing the GR point-like singularity can be found. The line element after solving the corresponding static, spherically symmetric field equations reads now as
\begin{equation} \label{eq:linelefR}
ds^2=-A(x)dt^2+\frac{1}{A(x)} \left(\frac{dx}{f_R} \right)^2 +r_c^2 z(x)^2 d\Omega^2 \ ,
\end{equation}
where $f_R\equiv df/dR$, and the radial function $r^2(x)=x^2/f_R$ is written for convenience as $r^2(x)=r_c^2 z^2(x)$, with $r_c$ containing constants coming from the matter sector. The metric functions take now highly involved expressions but, as we are just interested in their behavior around the minimum of the radial function, $z(x)_{min}=z_c$ (corresponding to the zeroes of $f_R$), we can bring here from \cite{bcor15,or15c} the relevant expansions at this minimum as
\begin{eqnarray}
z(x)&\approx& z_c + b(\alpha)x^2 \label{eq:zxfr} \\
f_R(z) &\approx& c(\alpha) (z-z_c) + \mathcal{O}(z-z_c)^2 \label{eq:fR} \\
A(z) &\approx& \frac{\delta_1 d(\alpha)}{(z-z_c)^2} + \mathcal{O}(z-z_c)^{-3/2} \ , \label{eq:AfR}
\end{eqnarray}
where $b(\alpha)>0, c(\alpha)>0$ and $d(\alpha)>0$ are some (matter model-dependent) constants, while $\delta_1$ encodes the relevant physical quantities: mass, charge and either the non-linear electromagnetic or anisotropic fluids characteristic parameters (see \cite{bcor15,or15c}, respectively, for details). For generic $f(R)$ models the geodesic equation can be written as
\begin{equation} \label{eq:geoeq}
\frac{ds}{dz} = \pm \frac{f_R^{1/2}\left(1+\frac{zf_{R,z}}{2f_R}\right)}{\sqrt{E^2 f_R^2 - A(z)f_R^2
\left(k+\frac{L^2}{z^2} \right)}} \ .
\end{equation}
For null radial geodesics ($k=0$ and $L=0$), the integration of this geodesic equation with the expressions (\ref{eq:fR}) and (\ref{eq:AfR}) shows that such geodesics take an infinite affine time $s$ to reach the wormhole throat. Consequently, the throat lies on the future (or past) boundary of the manifold. Moreover, geodesic timelike observers will find an infinite potential barrier before reaching the throat due to the divergence of $A(x)$ at $z=z_c$ and, consequently, will bounce off at some radius $z>z_c$. This means that no causal geodesic will be able to interact with the wormhole throat or with the universe {\it hidden} beyond it.

The pertinent question to ask now is the following: could an accelerated observer or particle go through the wormhole and get into the universe on the other side? Like in the EiBI case, the geometry (\ref{eq:linelefR}) can be cast under the form (\ref{eq:ley}) via the new radial coordinate $dy^2=dx^2/f_R^2$. This way, by noting that in this case $(u^y)^2=(u^x)^2/f_R^2=(dz/ds)^2/f_R$, the master equation (\ref{eq:acceL}) reads now\footnote{For simplicity, we assume here the same electric acceleration as in GR and in the EiBI gravity theory. Strictly speaking, the matter source that generates the geometry in the $f(R)$ case can be interpreted as a nonlinear theory of electrodynamics and, therefore, the relation between the Faraday tensor and the radial function $r(x)$ is different from the Maxwellian one. Since we are not specifying the form of the matter source and the acceleration is chosen for illustrative purposes, there is no need to use the precise form of the electric field that is consistent with this geometry.}
\begin{eqnarray}
\frac{1}{f_R}\left(\frac{dz}{ds}\right)^2&=&\left(E +\frac{Qq}{m r_c} \int_{z_0}^z \frac{dz'}{z^2\sqrt{f_R\left(1+\frac{L^2}{z^2(x)}\right)}} \right)^2 \nonumber \\
&-&A(z) \left(1+\frac{L^2}{z^2} \right) \ .
\end{eqnarray}
Focusing on the near-throat region, we can use the approximate expression (\ref{eq:fR}) to obtain
\begin{equation}
I^{f}_L\approx \frac{Qq}{m r_c} \frac{(K_0+2\sqrt{z-z_c})}{z_c^2\sqrt{c(\alpha)\left(1+\frac{L^2}{z_c^2}\right)}} \ ,
\end{equation}
where $K_0$ denotes an integration constant. Like in the EiBI case, we find that a bounded acceleration does not significantly alter the behavior of massive particles near the wormhole because the effect of the term $I^f_L$ boils down to a shift in the energy parameter $E$. In order to be able to reach and cross the wormhole throat, an unbounded acceleration would be necessary in order to overcome the infinite effective barrier represented by the term $A(z)\sim 1/(z-z_c)^2$. Thus we conclude that these geometries retain their nonsingular character regarding accelerated observers with bound acceleration. The unbound acceleration scenario is discussed in the Appendix for the two geometries of this section.

\section{Conclusions and discussion} \label{sec:V}

The notion of geodesic completeness stands at the root of the singularity theorems and is regarded as a key element to determine whether a spacetime is singular or not. The principle of general covariance, however, tells us that the laws of Physics need not be specified by any particular set of observers and that arbitrary reference frames are equally valid choices. For this reason, in this paper we have analyzed the trajectories of accelerated observers, since in any non-singular spacetime one should require the completeness of all timelike trajectories, regardless of whether they are freely falling or accelerated (non-gravitational interactions/accelerations should not spoil the good properties of the spacetime).

The specific scenario considered here is that provided by two different geodesically complete black hole spacetimes with wormhole structure. In the first example, which can be seen as an extension of the Reissner-Nordstr\"{o}m solution of GR to the case of Eddington-inspired Born-Infeld gravity, the wormhole is traversable by geodesic observers under some circumstances ($\delta_1\leq \delta_c$). Assuming an unbounded acceleration we have found that the situation is qualitatively identical to that observed for freely falling observers. This means that the trajectories of the particles have a unique solution on both sides of the wormhole and can be smoothly continued across the throat, thus confirming their completeness.

In the second case, a quadratic $f(R)$ theory is coupled to an anisotropic fluid (or, equivalently, to a nonlinear theory of electrodynamics) yielding a geodesically complete black hole spacetime  with wormhole structure in which the wormhole lies on a boundary of the manifold. This conclusion follows from the fact that massive particles find an infinite potential barrier  before reaching the throat, which makes them bounce, while null radial geodesics (light rays) take an infinite time to reach the throat. We have shown that the situation with accelerated observers is pretty much the same as in the EiBI case, in that timelike trajectories with bound accelerations bounce before reaching the wormhole. For the same reasons as in the Born-Infeld case, these accelerated trajectories are also complete.

An aspect that is worth emphasizing is that geodesic completeness is possible even though curvature scalars generically diverge at the wormhole throat in the two examples considered. These divergences may be regarded as irrelevant in the $f(R)$ case because they lie on a boundary of the manifold (beyond the reach of geodesics) but are certainly accessible in the Born-Infeld case, since the wormhole is traversable by geodesic observers in some cases. The infinite tidal forces caused by those divergences seem to have little effect on the geodesics and also on extended objects, as shown in \cite{ors16a}. In the Appendix we provide an example to illustrate that infinite non-gravitational accelerations do not necessarily have a pathological impact on timelike observers either, which is thus another piece of evidence supporting the non-singular character of the geometries considered here. Further confirmation of this may be obtained by extending the analysis carried out in \cite{ors16a} for congruences of geodesics to the case of accelerated extended bodies. Work in this direction is currently underway.

\section*{Acknowledgments}

G. J. O. is funded by a Ramon y Cajal contract No. RYC-2013-13019. D. R.-G. is funded by the Funda\c{c}\~ao para a Ci\^encia e a Tecnologia (FCT, Portugal) postdoctoral fellowship No.~SFRH/BPD/102958/2014 and the FCT research grant UID/FIS/04434/2013. This work is supported by the Spanish grant FIS2014-57387-C3-1-P (MINECO/FEDER, EU), the project H2020-MSCA-RISE-2017 Grant FunFiCO-777740, the project SEJI/2017/042 (Generalitat Valenciana), the Consolider Program CPANPHY-1205388, and the Severo Ochoa grant SEV-2014-0398 (Spain). This article is based upon work from COST Action CA15117, supported by COST (European Cooperation in Science and Technology). We thank Jos\'{e} M. M. Senovilla for bringing to our attention the need to avoid discrimination among time-like observers based on their acceleration, which gave rise to the idea behind this paper, and also for his useful comments on a previous version of it.

\section*{Appendix: Observers with unbounded acceleration}

For completeness of this work, let us extend our previous analysis to accelerated observers with unbound accelerations. This is motivated on the grounds that, if an unbounded acceleration yields complete trajectories, it seems natural to expect that bounded accelerations should also do it. Given that arbitrarily large accelerations maintained over an arbitrarily large period of time are not admissible from an energetic point of view, we limit our attention to accelerations which can be unbounded as certain regions are approached, such as those experienced by charged particles.

For the case of EiBI gravity in section (\ref{sec:IVA}), if instead of a standard electric field force, we propose a slightly modified version of the form $Qq/(r-r_c)^2$,  in order to generate an unbounded acceleration at the throat, one finds that (for the case $L=0$ and near the throat) Eq.(\ref{eq:beyondGeoBI}) should be replaced by
\begin{equation}\label{eq:beyondGeoBImod}
\left(\frac{dx}{2d\tau}\right)^2\approx \left(\tilde{E} -\frac{Qq}{3m r_c}\left(\frac{r_c}{r-r_c}\right)^{\frac{3}{2}} \right)^2 -A(x) \ .
\end{equation}
Now the first term on the right-hand side always diverges as  $\sim 1/|x|^6$ as the throat at $x=0$ is approached. In the case $\delta_1<\delta_c$, this divergent positive term  is added to the divergence of $A(x)\sim -1/|x|$ and helps the particle reach the wormhole faster than in the geodesic case (regardless of the sign of the charges!). If  $\delta_1>\delta_c$, then $A(x)\sim +1/|x|$ contributes to diminish the magnitude of the right-hand side and, depending on the particular choice of parameters, the right-hand side could remain positive all the way down to $x=0$, thus allowing accelerated particles to go through the wormhole. For relatively weak accelerations (small $Qq/m$), however, it is expected that the right-hand side vanishes at some point before reaching the throat, forcing the bounce of the particle in much the same way as in the freely falling case.

The lesson that follows from this artificial example is that particles acted upon by an unbounded force could reach regions which are forbidden to geodesic observers. In this particular model, the only requirement for this to happen is that in the $\delta_1>\delta_c$ case, the term $(\tilde{E}+\int^{y(x)} k(y')dy')^2$ should grow faster than $A(x)$ so that the subtraction of those two terms remains positive for some observers. In any case, since the integration of $s=s(x)$ until $x=0$ is continuous and unique, it can be smoothly extended across the wormhole, confirming that all such trajectories are complete, like in the geodesic scenario.

The results above hold also in the case of $f(R)$ theories of section (\ref{eq:subb}). Indeed, should any unbound acceleration in that case be physically possible, then a certain family of observers could be able to cross the wormhole throat but, since the curves $s=s(z)$ are continuous and unique even at $z=z_c$ (or $x=0$), nothing would prevent their extension across the throat, which confirms their completeness also in these cases.


\begin{thebibliography}{99}

\bibitem{Penrose1}
R. Penrose, Phys. Rev. Lett. \textbf{14}, 57 (1965);
Riv. Nuovo Cim. Numero Speciale \textbf{1}, 252 (1969);
Gen. Relativ. Gravit. \textbf{34}, 1141 (2002).
\bibitem{Hawking}
S. W. Hawking, Phys. Rev. Lett. \textbf{17}, 444 (1966).
\bibitem{Carter}
B. Carter, Phys. Rev. Lett. \textbf{26}, 331 (1971).
\bibitem{Senovilla1}
J.~M.~M.~Senovilla and D.~Garfinkle, Class.\ Quant.\ Grav.\  {\bf 32}, 124008 (2015);
J.~M.~M.~Senovilla, Gen. Rel. Grav. (1998) 30: 701.
\bibitem{Wald}
R.~M.~Wald, \emph{General Relativity}  (University of Chicago Press, Chicago, 1984).
\bibitem{Senovilla2}
J. M. Senovilla, Phys. Rev. Lett. \textbf{64}, 2219 (1990).
\bibitem{boundedscalars0}
S. Ansoldi, arXiv:0802.0330[gr-qc].
\bibitem{BS2}
E.~Ayon-Beato and A.~Garcia, Phys.\ Rev.\ Lett.\  {\bf 80},  5056 (1998);
I.~Dymnikova,  Class.\ Quant.\ Grav.\  {\bf 21}, 4417 (2004);
W.~Berej, J.~Matyjasek, D.~Tryniecki and M.~Woronowicz, Gen.\ Rel.\ Grav.\  {\bf 38},  885 (2006);
P.~Nicolini, A.~Smailagic and E.~Spallucci,  Phys.\ Lett.\ B {\bf 632},  547 (2006);
J.~P.~S.~Lemos and V.~T.~Zanchin, Phys.\ Rev.\ D {\bf 83},  124005 (2011);
C.~Rovelli and F.~Vidotto,  Int.\ J.\ Mod.\ Phys.\ D {\bf 23}, 1442026 (2014);
V.~P.~Frolov, Phys.\ Rev.\ D {\bf 94}, 104056 (2016);
I.~Dymnikova and E.~Galaktionov,  Class.\ Quant.\ Grav.\  {\bf 33}, 145010 (2016);
E.~Spallucci and A.~Smailagic, Int.\ J.\ Mod.\ Phys.\ D {\bf 26}, 1730013 (2017).
\bibitem{ES}
G. F. R. Ellis and B. G. Schmidt, Gen. Rel. Grav. \textbf{8}, 915 (1977).
\bibitem{Tipler1}
F. J. Tipler, Phys. Rev. D \textbf{15}, 942 (1977);
Phys. Lett. A \textbf{64}, 8 (1977);
F. J. Tipler, C. J. S. Clarke, and G. F. R. Ellis, \emph{General Relativity and Gravitation} (Plenum, New York, 1980).
\bibitem{CK1}
A. Krolak, Class. Quant. Grav. \textbf{3}, 267 (1986);
C. J. S. Clarke and A. Krolak, J. Geom. Phys. \textbf{2}, 127 (1985).
\bibitem{Nolan}
B. C. Nolan, Phys. Rev. D {\bf 60}, 024014 (1999);
{\bf 62}, 044015 (2000).
\bibitem{Geroch68}
R. Geroch, Ann. Phys. \textbf{48}, 526 (1968).
\bibitem{Tod94}
K. P. Tod,  Class. Quant. Grav. \textbf{11}, 1331 (1994).
\bibitem{GL93}
D.~V.~Galtsov and P.~S.~Letelier, Phys.\ Rev.\ D {\bf 47},  4273 (1993).
\bibitem{Burko}
L.~M.~Burko and G.~Khanna, arXiv:1709.10155 [gr-qc].
\bibitem{Senovilla98}
J. M. M. Senovilla, Gen. Rel. Grav. \textbf{30}, 701 (1998).
\bibitem{Beem80}
J. K. Beem, Commun. Math. Phys. \textbf{72}, 273 (1980) .
\bibitem{Rin}
W. Rindler, Phys. Rev. \textbf{119}, 2082 (1960);
B. Mashhoon and U. Muench, Ann. Phys. \textbf{11}, 532 (2002);
E. Gourgoulhom, \emph{Special relativity in general frames} (Graduate Texts in Physics, Springer-Verlag, 2013).
\bibitem{SF16a}
T. Scarr and Y. Friedman, Gen. Rel. Grav. \textbf{48}, 65 (2016).
\bibitem{FR15a}
D. de la Fuente and A. Romero, Gen. Rel. Grav. \textbf{47}, 33 (2015).
\bibitem{ors15b}
G. J. Olmo, D. Rubiera-Garcia, and A. Sanchez-Puente, Eur. Phys. J. C \textbf{76}, 143 (2016).
\bibitem{blor17a}
J. Beltr\'an Jim\'enez, L. Heisenberg, G. J. Olmo, and D. Rubiera-Garcia, Physics Reports, in press, arXiv:1704.03351 [gr-qc].
\bibitem{ors15a}
G. J. Olmo, D. Rubiera-Garcia, and A. Sanchez-Puente, Phys. Rev. D \textbf{92}, 044047 (2015).
\bibitem{ors16a}
 G. J. Olmo, D. Rubiera-Garcia, and A. Sanchez-Puente, Class. Quant. Grav. \textbf{33}, 115007 (2016).
\bibitem{SF16b}
Y. Friedman and T. Scarr, Gen. Rel. Grav. \textbf{47}, 121 (2015).
\bibitem{OlmoBook}
G.~J.~Olmo, Springer Proc.\ Phys.\  {\bf 176}, 183 (2016).
\bibitem{DG}
S. Deser and G. W. Gibbons, Class. Quant. Grav. {\bf 15}, L35 (1998).
\bibitem{BIapp}
M. Ba\~nados, Phys. Rev. D \textbf{77}, 123534 (2008);
M. Ba\~nados, P. G. Ferreira, and C. Skordis, Phys. Rev. D \textbf{79}, 063511 (2009);
P. P. Avelino and R. Z. Ferreira, Phys. Rev. D \textbf{86}, 041501 (2012);
P. Pani, T. Delsate, and V. Cardoso, Phys. Rev. D \textbf{85} (2012) 084020;
F. Fiorini, Phys. Rev. Lett. \textbf{111}, 041104 (2013);
M.~Bouhmadi-Lopez, C.~Y.~Chen, and P.~Chen, Eur.\ Phys.\ J.\ C {\bf 74}, 2802 (2014);
{\bf 75}, 90 (2015);
Phys.\ Rev.\ D {\bf 90},  123518 (2014);
C.~Y.~Chen, M.~Bouhmadi-Lopez, and P.~Chen, Eur.\ Phys.\ J.\ C {\bf 76}, 40 (2016);
S.~W.~Wei, K.~Yang and Y.~X.~Liu, Eur.\ Phys.\ J.\ C {\bf 75},  253 (2015);
S.~Jana and S.~Kar,  Phys.\ Rev.\ D {\bf 92},  084004 (2015);
{\bf 96},  024050 (2017);
R.~Shaikh, Phys.\ Rev.\ D {\bf 92},  024015 (2015);
P.~P.~Avelino, Phys.\ Rev.\ D {\bf 93}, 044067 (2016); 104054 (2016);
K.~Yang, Y.~X.~Liu, B.~Guo and X.~L.~Du, Phys.\ Rev.\ D {\bf 96}, 064039 (2017);
S.~L.~Li and H.~Wei, Phys.\ Rev.\ D {\bf 96}, 023531 (2017).
\bibitem{ors13}
G. J. Olmo, D. Rubiera-Garcia, and H. Sanchis-Alepuz, Eur. Phys. J. C \textbf{74}, 2804 (2014).
\bibitem{Visser}
M. Visser, \textit{Lorentzian Wormholes} (AIP Press, New York, 1996).
\bibitem{Lobo:2017oab}
F.~S.~N.~Lobo (Editor), \textit{Wormholes, Warp Drives and Energy Conditions}, Fundam.\ Theor.\ Phys.\  {\bf 189}, (2017).
\bibitem{Chandra}
S. Chandrasekhar, \emph{The mathematical theory of black holes} (Oxford University Press, New York, 1992).
\bibitem{fRr}
A. De Felice and S. Tsujikawa, Liv. Rev. Rel. \textbf{13}, 3 (2010);
S. Capozziello and M. De Laurentis, Phys. Rep. \textbf{509}, 167 (2011);
S. Nojiri and S. D. Odintsov, Phys. Rep. \textbf{505}, 59 (2011);
G. J. Olmo, Int.\ J.\ Mod.\ Phys.\  D {\bf 20}, 413 (2011).
\bibitem{fR}
S.~Nojiri, S.~D.~Odintsov, and D.~Saez-Gomez, Phys.\ Lett.\ B {\bf 681},  74 (2009);
S.~H.~Hendi and D.~Momeni, Eur.\ Phys.\ J.\ C {\bf 71},  1823 (2011);
T.~Moon, Y.~S.~Myung and E.~J.~Son, Gen.\ Rel.\ Grav.\  {\bf 43},  3079 (2011);
A.~V.~Astashenok, S.~Capozziello, and S.~D.~Odintsov,  JCAP {\bf 1312},  040 (2013).
\bibitem{bcor15}
C. Bambi, A. Cardenas-Avendano, G. J. Olmo, and D. Rubiera-Garcia,  Phys. Rev. D \textbf{93}, 064016 (2016).
\bibitem{or15c}
G. J. Olmo and D. Rubiera-Garcia, Universe {\bf 1}, 173 (2015);
C.~Bejarano, G.~J.~Olmo, and D.~Rubiera-Garcia,  Phys.\ Rev.\ D {\bf 95}, 064043 (2017).



\end{thebibliography}
\end{document}